\documentclass{article}
\usepackage{amssymb}
\usepackage{amsmath}
\usepackage{epsfig,graphics}

\newcommand{\bea}{\begin{eqnarray*}}
\newcommand{\eea}{\end{eqnarray*}}
\newcommand{\alfa}{\alpha}
\newcommand{\teta}{\theta}
\newcommand{\ro}{\rho}

\newcommand{\som}{\mbox{sum}}

\begin{document}

\title{Scaling a unitary matrix}
\author{Alexis De Vos$^1$ and 
        Stijn De Baerdemacker$^2$\footnote{SDB is an `FWO-Vlaanderen' post-doctoral fellow
                                           and thanks Dimitri Van Neck and Brecht Verstichel of UGent
                                           for valuable discussions.} \\
        $^1$ Cmst, Imec v.z.w. and vakgroep elektronica en informatiesystemen, \\ 
        Universiteit Gent, B - 9000 Gent, Belgium \\
        {\tt alex@elis.ugent.be}  \\
        $^2$ Center for Molecular Modeling, \\
        vakgroep fysica en sterrenkunde, \\ 
        Universiteit Gent, B - 9000 Gent, Belgium \\
        {\tt Stijn.DeBaerdemacker@ugent.be}  
        }
\date{$\ $} 

\maketitle

\begin{abstract}
The iterative method of Sinkhorn allows, 
starting from an arbitrary real matrix with non-negative entries,
to find a so-called `scaled matrix' which is doubly stochastic,
i.e.\ a matrix with all entries in the interval (0,~1) and with all line sums equal to~1.
We conjecture that a similar procedure exists, which allows,
starting from an arbitrary unitary matrix, to find a scaled matrix
which is unitary and has all line sums equal to~1.
The existence of such algorithm guarantees a powerful decomposition
of an arbitrary quantum circuit.
\end{abstract}

\section{Introduction}

By definition, the scaling of an $n \times m$ matrix $A$ is the multiplication of this matrix 
to the  left by a diagonal $n \times n$ matrix $L$ and 
to the right by a diagonal $m \times m$ matrix $R$,
resulting in the $n \times m$ matrix $B = LAR$, called the scaled matrix \cite{amestoy}.

Sinkhorn \cite{sinkhorn} has demonstrated that an arbitrary matrix~$A$ 
with exclusively real and positive entries 
can be scaled by diagonal matrices $L$ and $R$ with exclusively positive real entries,
such that the resulting scaled matrix~$B$ is doubly stochastic,
i.e.\ such that all entries of $B$ are real and in the interval $(0, 1)$ and all line sums
(i.e.\ all row sums and all column sums) of $B$ are equal to~1. 

In order to find the appropriate matrices $L$ and $R$, 
one could consider the set of $n+m$ line-sum equations
and solve it for the $n$ unknown entries of $L$ and the $m$ unknown entries of $R$.
However, all equations are quadratic. 
As the equations are non-linear, an analytic solution of the set is not available. 
Instead, one proceeds iteratively,
in the way pioneered by Kruithof \cite{kruithof}.
One computes the matrices $A_1$, $A_2$, ..., successive approximations of the wanted matrix~$B$.
At the $k$~th iteration,
one chooses a diagonal matrix $L_k$ with diagonal entries equal to the inverse of the row sums
of the matrix $A_{k-1}$ :
\[
(L_k)_{aa} = 1\, /\, \sum_{b=1}^m (A_{k-1})_{ab}
\]
and one chooses a diagonal matrix $R_k$ with diagonal entries 
equal to the inverse of the resulting column sums
\[
(R_k)_{bb} = 1\, /\, \sum_{a=1}^n (L_kA_{k-1})_{ab} \ ,
\]
thus leading to the new matrix
\[
A_k = L_k A_{k-1} R_k\ .
\]
Because this procedure converges, the matrices $L_k$ and $R_k$ ultimately become the 
$n \times n$ and $m \times m$ unit matrices, respectively.
This means that ultimately $A_k$ becomes a matrix with unit line sums.
By choosing $A_0$ equal to $A$, 
the two wanted scaling matrices are
\bea
L & = & L_{\infty} ... L_2 L_1 \\
R & = & R_1 R_2 ... R_{\infty} 
\eea
and the scaled matrix $B$ is $A_{\infty}$.

In the present paper, we investigate whether it is possible
to scale an $n \times n$ unitary matrix $A$ by two unitary diagonal matrices $L$ and $R$,
such that the scaled matrix $B=LAR$ has all line sums equal to~1.
For this purpose, we apply a Sinkhorn-like algorithm, however with
\bea
(L_k)_{aa} & = & 1\, /\, \Phi\left(\ \sum_{b=1}^n    (A_{k-1})_{ab}\ \right)  \\[1mm]
(R_k)_{bb} & = & 1\, /\, \Phi\left(\ \sum_{a=1}^n (L_kA_{k-1})_{ab}\ \right)  \ ,
\eea
thus guaranteeing that all the diagonal entries of $L_k$ and $R_k$
are automatically unitary.
Here, we call a complex number~$x$ unitary iff $|x|=1$
and define the function $\Phi$ of a complex number~$y$ as
\bea
\Phi (y) & = & \frac{y}{|y|} = \exp(i\, \arg(y)) \hspace{5mm}    \mbox{ if } |y|>0 \\[1mm]
         & = & 1                                 \hspace{32.5mm} \mbox{ if } |y|=0 \ .
\eea
If this procedure ultimately converges, then both $L_k$ and $R_k$ ultimately become the 
$n \times n$ unit matrix.
This means that ultimately $A_k$ becomes a matrix with all line sums having $\Phi=1$,
i.e.\ with all line sums real (positive or zero).
Below, we will see that not only $A_k$ converges to a matrix $A_{\infty}$
with exclusively 
real\footnote{The $n \times n$ unitary matrices with real line-sums form a subset of U($n$),
              but not a subgroup of U($n$). This can easily be illustrated by multiplying
              two U(2) matrices: the square root of {\tt NOT} matrix 
              $M_1 = \frac{1}{2}\ {\scriptsize \left( \begin{array}{rr}      1+i & 1-i \\ 1-i &      1+i \end{array} \right)}$
              and the orthogonal matrix  
              $M_2 = \frac{1}{2}\ {\scriptsize \left( \begin{array}{rr} \sqrt{3} & -1  \\ 1   & \sqrt{3} \end{array} \right)}$.
              All four line sums of both matrices are real (and positive); 
              their product~$M_1M_2$, however, has 
              a real and positive first column sum $c_1=(\sqrt{3}+1)/2$, but 
              a complex           first row    sum $r_1=(\sqrt{3}-i)/2$.} 
(non-negative) line sums,
but that surprisingly those line sums moreover equal~1.

\section{A progress measure}

For investigating the progress in the matrix sequence $A_0, A_1, A_2, ...$, 
we basicly follow a reasoning similar to
the elegant proofs of Sinkhorn's theorem 
by Linial et al.\ \cite{linial} \cite{wigderson} and 
by Aaronson \cite{aaronson}.
However, the pivotal role (called either `progress measure' or `potential function') 
played either 
by the matrix permanent (Linial et al.) or
by the matrix product   (Aaronson) 
is taken over here by the absolute value of the matrix sum.
The matrix sum of an $n \times m$ matrix $X$ is defined as the sum of all its entries: 
\[
\som(X) = \sum_{b=1}^m \sum_{a=1}^n X_{ab} \ .
\]

Assume a matrix $X$ with row sums $r_a$ and column sums $c_b$.
We denote by $L$ the diagonal matrix with entries $L_{aa}$ equal to $1/\Phi(r_a)$.
If $X'$ is a short-hand notation for $LX$, 
and $r'_a$ and $c'_b$ are its row sums and column sums, respectively, then
we have
\begin{equation}
|\, \som(X')\, | = |\,  \sum_{a=1}^n r'_a\, | \ge  |\, \sum_{a=1}^n r_a\, |  = |\, \som(X)\, | \ ,
\label{'}
\end{equation}
because $\sum_{a} r'_a$ can be regarded as a vector sum of vectors 
with the same length as the vectors $r_a$ but with zero angles between them.
The equality sign holds iff all numbers $r_a$ have the same argument.
Similarly, we denote by $R$ the diagonal matrix with entries $R_{bb}$ equal to $1/\Phi(c'_b)$.
If $X''$ is a short-hand notation for $X'R$, and $c''_b$ are its column sums, then   
\begin{equation}
|\, \som(X'')\, | = |\, \sum_{b=1}^m c''_b\ | \ge  |\, \sum_{b=1}^m c'_b\, |  = |\, \som(X')\, | \ ,
\label{''}
\end{equation}
where the equality sign holds iff all numbers $c'_b$ have the same argument.
We thus can conclude that
\[
|\, \som(X'')\, | \ge |\, \som(X)\, | \ ,
\]
where the equality holds iff the equality holds both in (\ref{'}) and in (\ref{''}).
The equality in~(\ref{'})  occurs if all $r_a$  have the same argument, whereas
the equality in~(\ref{''}) occurs if all $c'_b$ have the same argument.
But, a constant $\arg(r_a)$ leads to an $X'$ of the form $e^{i\alfa}\, X$ and therefore
to column sums $c'_b = e^{i\alfa}\, c_b$ and 
the condition of constant $\arg(c'_b)$ then is equivalent to 
the condition of constant $\arg(c_b)$.

We conclude that, in the matrix sequence $A_0, A_1, A_2, ...$ of Section~1, we have  
\begin{equation}
|\, \som(A_k)\, | \ge |\, \som(A_{k-1})\, | \ ,
\label{||}
\end{equation}
where the equality sign holds
iff $A_{k-1}$ is a matrix with all line sum arguments equal.
As soon as $k-1>0$, this condition is equivalent to
iff $A_{k-1}$ is a matrix with all line sums real, either zero or 
positive\footnote{We remark that, for $k > 0$, 
                  the procedure of Section~1 guarantees that 
                  all column sums and thus also the matrix sum 
                  are positive or zero. 
                  Hence, for $k \ge 2$, the absolute value symbols in (\ref{||}) could be omitted.}. 

In Appendix~A, we prove that, for an arbitrary $n \times n$ unitary matrix~$U$, 
we have $|\som(U)| \le n$.
The equality sign holds iff $U$ is, up to a global phase, a matrix with all line sums equal to~1.
For sake of convenience, we define the potential function~$\Psi$ of an $n \times n$ matrix~$M$ as
\[
\Psi(M) = n^2 - |\som(M)|^2\ ,
\]
such that for all unitary matrices $0 \le \Psi \le n^2$ holds and the zero potential 
corresponds with unit line-sum matrices (times a factor $e^{i\alfa}$).
With this convention, we rewrite (\ref{||}) as 
\[
\Psi(A_k) \le \Psi(A_{k-1})\ .
\] 

The $\Psi$ landscape of the matrix group U($n$) displays stationary points.
In Appendix~B, we show that these points 
either have zero matrix sum 
    or have all (non-zero) line sums with same argument.
We distinguish three categories of stationary points: their potential satisfies 
\begin{itemize}
\item either $\Psi = n^2$,
\item     or $\Psi = 0$,
\item     or $0 < \Psi < n^2$.
\end{itemize}
If the first  case occurs, then the stationary point is a global maximum;
if the second case occurs, then the stationary point is a global minimum.
We conjecture that the third class consists of saddle points.
In other words: 
we conjecture that the $\Psi$ landscape has 
no local minima (nor local maxima).
As a result, the scaling procedure 
(with ever decreasing $\Psi$) ultimaltely converges to 
the point with minimal potential.
We conjecture that this global minimum is a matrix with $\Psi=0$ 
and thus is a wanted unit line-sum matrix~$B$.

We distinguish two cases:
\begin{itemize}
\item Either the given matrix $A$ does not have constant line sum arguments,
      in which case we choose $A_0=A$. As long as the subsequent matrices 
      $A_1$, $A_2$, ... do not have all row sums real,
      we have a strictly decreasing sequence 
      $\Psi(A_0) > \Psi(A_1) > \Psi(A_2) > ...$. 
      This sequence is bounded by~0.
      Therefore a limit matrix $A_{\infty}$ exists, with $\Psi(A_{\infty}) \ge 0$.
      \begin{itemize}
      \item If $\Psi(A_{\infty}) = 0$, then $A_{\infty}$ is the wanted scaled matrix.
            The scaling matrices are $L= L_{\infty} ... L_2 L_1$ and $R = R_1 R_2 ... R_{\infty}$.
      \item In the (very unlikely) case that $\Psi(A_{\infty}) > 0$, 
            the matrix $A_{\infty}$ is a stationary point in the potential landscape~$\Psi$.
            According to the conjecture, this point is a saddle point and therefore
            we can apply appropriate matrices $L$ and $R$, both close to the unit matrix,
            such that $LA_{\infty}R$ has potential $\Psi$ lower than $\Psi(A_{\infty})$.
            It is sufficient to try $n$~mutually orthogonal directions
            in the $(2n-1)$-dimensional neighbourhood of the saddle point.
            After applying these $L$ and $R$,
            we restart the algorithm with a new $A_0$, equal to $LA_{\infty}R$.
      \end{itemize}
\item Or the given matrix~$A$ has all line sum arguments equal and thus $A$ is a stationary point.
      In this case we choose $A_0=L_0AR_0$, with two appropriate matrices $L_0$ and $R_0$,
      such that the start matrix $A_0$ is not a stationary point. 
      For this purpose, we can proceed as follows:
      \begin{itemize}
      \item If at least two row sums of $A$ are different from~0,
            e.g.\ $r_{x} \neq 0$ and $r_{y} \neq 0$,
            then we take $R_0$ equal to the unit matrix and all entries of $L_0$ equal to~1,
            except $(L_0)_{xx}$, 
            thus resulting in at least two different row-sum arguments for $A_0$.
      \item If at least two column sums of $A$ are different from~0,
            e.g.\ $c_{x} \neq 0$ and $c_{y} \neq 0$,
            then we take $L_0$ equal to the unit matrix and all entries of $R_0$ equal to~1,
            except $(R_0)_{xx}$, 
            thus resulting in at least two different column-sum arguments for $A_0$.
      \item If only one row sum (say $r_x$) and only one column sum (say $c_y$) of $A$ differ from~0,
            i.e.\ if $A$ is a generalized Hadamard matrix \cite{tadej},
            then we take $R_0$ equal to the unit matrix and all entries of $L_0$ equal to~1,
            except $(L_0)_{xx}$, 
            thus resulting in at least two different column-sum arguments for $A_0$.
      \end{itemize}
\end{itemize}
An example of the latter case is the orthogonal matrix
\[
A = \left( \begin{array}{rr}  \cos(\phi) & \sin(\phi) \\  
                             -\sin(\phi) & \cos(\phi) \end{array} \right) \ ,
\]
where, for convenience, we assume $0 < \phi < \pi/4$. 
Indeed: all its line sums have zero argument.
If we would take $A_0=A$, then
$L_1$ would be equal to the $2 \times 2$ unit matrix and subsequently 
$R_1$ would be equal to the              unit matrix,                
such that $A_1$ and, in fact, all subsequent $A_k$ would be equal to $A$
and therefore $\Psi(A_0) = \Psi(A_1) = \Psi(A_2) = ...$,
equal to $2\cos(\phi)$ in this example.
In this way, the algorithm cannot find the wanted solution,
in spite of the fact that such scaled matrices with exclusively unit line sums actually exist,
e.g.\  
\[
\hspace*{-8mm}
B = \left( \begin{array}{cc}    e^{i\phi} & 0 \\ 0 &  i\, e^{i\phi} \end{array} \right) \, A\  
    \left( \begin{array}{rr}   1          & 0 \\ 0 & -i             \end{array} \right)
  = \left( \begin{array}{rr}   \cos(\phi)\, e^{i\phi} & -i\sin(\phi)\, e^{i\phi} \\ 
                             -i\sin(\phi)\, e^{i\phi} &   \cos(\phi)\, e^{i\phi} \end{array} \right) \ .
\]
In order to avoid the no-start of the convergence towards the desired scaled matrix~$B$,
we apply e.g.\ the matrices 
$L_0 = {\scriptsize \left( \begin{array}{rr} i & 0 \\ 0 & 1 \end{array} \right)}$ and
$R_0 = {\scriptsize \left( \begin{array}{rr} 1 & 0 \\ 0 & 1 \end{array} \right)}$,
resulting in 
\[
A_0 = L_0AR_0 = \left( \begin{array}{rr} i\cos(\phi) & i\sin(\phi) \\  
                                         -\sin(\phi) &  \cos(\phi) \end{array} \right) \ ,
\]
where row sums $r_1 = i(\cos(\phi)+\sin(\phi))$ and $r_2 = \cos(\phi)-\sin(\phi)$ 
indeed have unequal arguments: $\pi/2$ and 0, respectively.

\section{Convergence speed}

As a first example of the procedure of Section~2, we take the unitary matrix
\[
A = \frac{1}{4}\ \left( \begin{array}{ccc}  1    & 1-3i & -2+i \\  
                                           -1-3i & 2    &  1+i \\
                                            2+ i & 1- i &  3   \end{array} \right) \ ,
\]
with line sums, matrix sum, and potential
\bea
r_1  & = &    - i /2 \\
r_2  & = & (1 - i)/2 \\
r_3  & = &  3     /2 \\
c_1  & = & (1 - i)/2 \\
c_2  & = &  1 - i    \\
c_3  & = & (1 + i)/2 \\
m    & = &  2 - i    \\
\Psi & = &  4 \ .
\eea
Because $\Psi \neq n^2$, $\Psi \neq 0$, and the line sums do not have equal argument,
we are in a `common' case, i.e.\  not in a stationary point of the $\Psi$ landscape.
We thus choose $A_0=A$. 
Subsequent steps of the algorithm yield potentials
\bea
\Psi(A_0) & = & 4.00000 \\
\Psi(A_1) & = & 1.16956 \\
\Psi(A_2) & = & 0.44189 \\
\Psi(A_3) & = & 0.17167 \\
\Psi(A_4) & = & 0.07723 \\
\Psi(A_5) & = & 0.03885 \ ,
\eea
approximately decreasing exponentially.

E.g.\ for $k=5$, we have
\[
A_5 = \left( \begin{array}{rrr} -0.2398 + 0.0708\, i &  0.7522 + 0.2432\, i &  0.4337 - 0.3527\, i \\
                                 0.7113 - 0.3451\, i &  0.4945 + 0.0739\, i & -0.2341 + 0.2649\, i \\
                                 0.4871 + 0.2742\, i & -0.1564 - 0.3171\, i &  0.7448 + 0.0878\, i \end{array} \right) \ ,
\]
with line sums indeed close to~1:
\bea
r_1 & = & 0.9462 - 0.0386\, i \\
r_2 & = & 0.9717 - 0.0062\, i \\
r_3 & = & 1.0756 + 0.0449\, i \\
c_1 & = & 0.9587              \\
c_2 & = & 1.0904              \\
c_3 & = & 0.9444              \ .
\eea

Next, with the method of \.{Z}yczkowski et al.\ \cite{zyczkowski} \cite{pozniak}, 
we generate 1,000 random elements of~U(3),
uniformly distributed with respect to the Haar measure.
Table \ref{tabelletje} shows how the potential $\Psi$ 
decreases after each step of the scaling procedure.
Whereas the $\Psi$ values of the initial matrices~$A_0$ 
have a wide distribution between~0 and~$n^2=9$,
the distribution of $\Psi(A_k)$ is very peaked at $\Psi=0$, as soon as $k>0$.

The convergence speed turns out to be strongly different for different matrices~$A$.
Among the 1,000 samples, some converge exceptionally slowly, 
as is illustrated by the column `maximum($\Psi$)'.
Usually, however, convergence is fast,
as is illustrated by the column `average($\Psi$)'.
We stress the fact that all 1,000 experiments directly converge to the global minimum~$\Psi=0$,
and thus none `gets trapped' in a local minimum and none temporarily `halts' in a saddle point.

Finally, similar experiments 
with 1,000 random elements from~U(4) (see Table~\ref{tabelletje}) and 
with 1,000 random elements from~U(5)
lead to similar results.

\begin{table}[t]
   \centering
   \caption{The probability distribution of the potentials $\Psi$ of 
            1,000 random $n \times n$ unitary matrices, 
            after $k$~steps in the numerical algorithm.}
   \vspace*{3mm}
   \begin{tabular}{|r|rrr|rrr|}
   \hline
   &&&&&& \\
            &                  &   $n=3$     &             &                  &   $n=4$     &             \\ 
   &&&&&& \\
   \hline
   &&&&&& \\
   $k$ $\ $ & $\ $ min($\Psi$) & ave($\Psi$) & max($\Psi$) & $\ $ min($\Psi$) & ave($\Psi$) & max($\Psi$) \\ 
   &&&&&& \\
   \hline
   &&&&&& \\
   0 $\ $ & 0.35382 & 5.86125 & 8.99799 & 0.99521 & 11.91981 & 15.99986 \\ 
   1 $\ $ & 0.00002 & 0.46696 & 2.98141 & 0.00165 &  1.04461 &  5.19014 \\ 
   2 $\ $ & 0.00000 & 0.22149 & 2.16004 & 0.00013 &  0.47963 &  3.59764 \\
   3 $\ $ & 0.00000 & 0.13886 & 1.81920 & 0.00003 &  0.31220 &  2.96001 \\
   4 $\ $ & 0.00000 & 0.09704 & 1.25805 & 0.00000 &  0.23075 &  2.47531 \\
   5 $\ $ & 0.00000 & 0.07311 & 1.15923 & 0.00000 &  0.18143 &  2.08629 \\
 \vdots $\ \ $ & & & & & &  \\[1mm]
  10 $\ $ & 0.00000 & 0.02973 & 0.75787 & 0.00000 &  0.08351 &  1.07794 \\ 
  20 $\ $ & 0.00000 & 0.01003 & 0.26781 & 0.00000 &  0.03917 &  1.06910 \\ 
  30 $\ $ & 0.00000 & 0.00621 & 0.26199 & 0.00000 &  0.02469 &  1.06418 \\ 
  40 $\ $ & 0.00000 & 0.00465 & 0.26085 & 0.00000 &  0.01702 &  1.04939 \\ 
  50 $\ $ & 0.00000 & 0.00371 & 0.26061 & 0.00000 &  0.01287 &  0.98529 \\ 
\vdots $\ \ $  & & & & & &  \\[1mm]
 100 $\ $ & 0.00000 & 0.00194 & 0.26054 & 0.00000 &  0.00712 &  0.97675 \\ 
   &&&&&& \\
   \hline
   \end{tabular}
   \label{tabelletje}
\end{table}

\begin{figure}
              \begin{center}
              \includegraphics[clip, width=120mm]{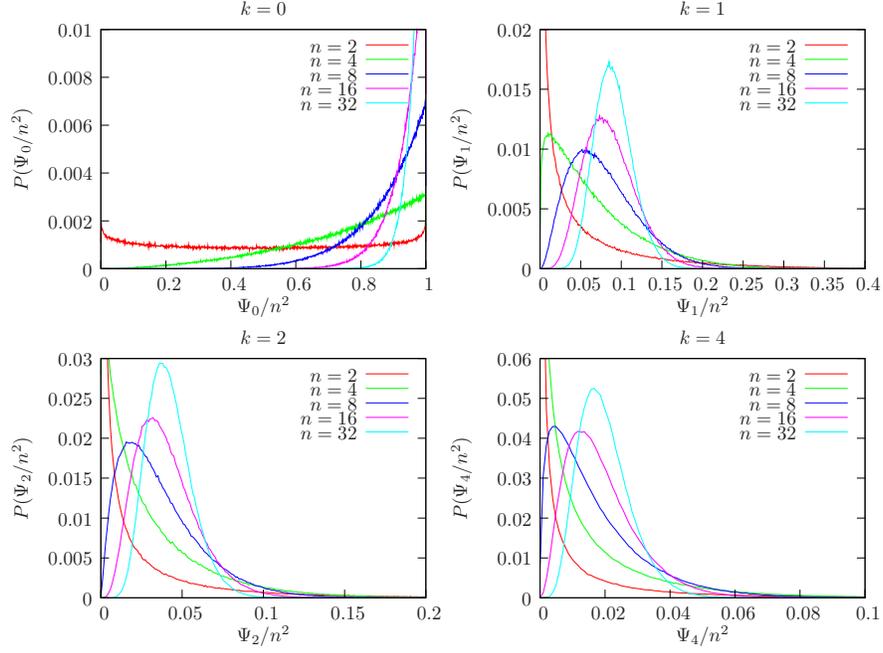}
              \caption{Probability distribution of the potential~$\Psi$
                       of $2^{20}$~random $n \times n$ unitary matrices, 
                       after~$k$~steps in its scaling procedure: 
                       (a)~after    0~steps,
                       (b)~after  one step,
                       (c)~after  two steps, and
                       (d)~after four steps.}
              \label{figuur1}
              \end{center}
\end{figure}

For $n$ equal to 2, 4, 8, 16, and 32, Figure~\ref{figuur1}
shows the probability distribution of the potential~$\Psi$,
after $k=0$, $1$, $2$, and $4$ iteration steps.
We see how the distribution, at each step,
shifts more and more to $\Psi=0$. 

\begin{figure}
              \begin{center}
              \includegraphics[clip, width=120mm]{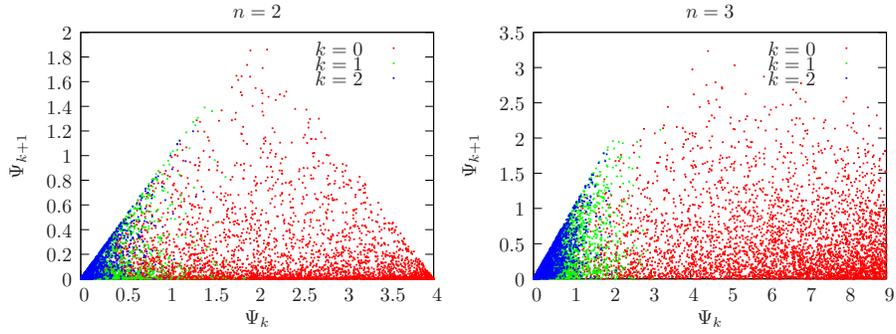}
              \caption{The    potential $\Psi$ after  an iteration step 
                       as a function of $\Psi$ before the          step:
                       $\Psi_1(\Psi_0)$   (red), 
                       $\Psi_2(\Psi_1)$ (green), and 
                       $\Psi_3(\Psi_2)$  (blue), for $2^{12}$~random matrices
                       (a)~from U(2) and
                       (b)~from U(3).}
              \label{figuur2}
              \end{center}
\end{figure}

The convergence can also be visualized by
displaying $\Psi_{k+1}$ as a function of $\Psi_{k}$,
i.e.\ the correlation between a $\Psi$ after and before an iteration step.
Figure~\ref{figuur2} shows 
$\Psi_1(\Psi_0)$, 
$\Psi_2(\Psi_1)$, and 
$\Psi_3(\Psi_2)$
for both $n=2$ and $n=3$.
As expected, all points lay below the line $\Psi_{k+1}=\Psi_k$. 
We see how the cloud of points, after each step, 
becomes smaller and smaller and moves closer and closer to 
the point $\Psi_{k} = \Psi_{k+1} = 0$.

\section{Application}

In Reference \cite{acm},
two subgroups of the unitary group U($n$) are presented:
\begin{itemize}
\item the subgroup ZU($n$), consisting of all $n \times n$ diagonal matrices
      with upper-left entry equal to~1 and other diagonal entries from U(1);
\item the subgroup XU($n$), consisting of all $n \times n$ unitary matrices with
      all of their $2n$ line sums are equal to~1,
\end{itemize}
and the following theorem is proved:
any U($n$) matrix~$U$ can be decomposed as 
\[
U = e^{i\alfa}\, Z_1X_1Z_2X_2Z_3...Z_{p-1}X_{p-1}Z_p \ ,
\]
with $p \le n(n-1)/2+1$ and where 
all $Z_j$ are ZU($n$) matrices and 
all $X_j$ are XU($n$) matrices.
In Reference \cite{bremen},
it is proved that a shorter decomposition exists: with $p \le n$.

In the present paper, we conjecture that even a far stronger theorem holds: $p \le 2$.
This means that
any U($n$) matrix~$U$ can be decomposed as 
\begin{equation}
U = e^{i\alfa}\, Z_1XZ_2 \ ,
\label{ZXZ}
\end{equation}
where both $Z_1$ and $Z_2$ are ZU($n$) matrices and $X$ is an~XU($n$) matrix.
In \cite{osid}, it is proved that the group
XU($n$) is isomorphic to U($n-1$) and hence has dimension $(n-1)^2$.
Therefore, the product $e^{i\alfa}\, Z_1XZ_2$ has
\begin{equation}
1 + (n-1) + (n-1)^2 + (n-1) = n^2
\label{dim}
\end{equation}
degrees of freedom,
matching exactly the dimension of U($n$) and thus making the conjecture 
dimensionally possible.
However, no analytic expression is provided for the unknown entries neither of
the matrices $Z_1$ and $Z_2$ nor of the scaled matrix~$X$.
An analytic solution of the decomposition problem is easily found for $n=2$. 
Indeed, an arbitrary member of U(2) can be decomposed 
according to (\ref{ZXZ}), in two different ways:
\[
e^{i\teta}\ 
\left( \begin{array}{cc}
       \cos(\phi)\, e^{ i\psi} & \sin(\phi)\, e^{ i\chi} \\
      -\sin(\phi)\, e^{-i\chi} & \cos(\phi)\, e^{-i\psi}  
       \end{array} \right) 
\]
\[ 
= e^{i\teta + i\phi + i\psi}\ 
\left( \begin{array}{cc}
       1 & 0 \\
       0 & i\, e^{-i\psi - i\chi} 
       \end{array} \right) \!\! 
\left( \begin{array}{rr}
        \cos(\phi)\, e^{-i\phi} & i\sin(\phi)\, e^{-i\phi} \\
       i\sin(\phi)\, e^{-i\phi} &  \cos(\phi)\, e^{-i\phi}  
       \end{array} \right) \!\!
\left( \begin{array}{cc}
       1 & 0 \\
       0 & -i\, e^{-i\psi + i\chi}  
       \end{array} \right)  
\] 
\[ 
= e^{i\teta - i\phi + i\psi}\ 
\left( \begin{array}{cc}
       1 & 0 \\
       0 & -i\, e^{-i\psi - i\chi} 
       \end{array} \right) \!\! 
\left( \begin{array}{rr}
         \cos(\phi)\, e^{i\phi} & -i\sin(\phi)\, e^{i\phi} \\
       -i\sin(\phi)\, e^{i\phi} &   \cos(\phi)\, e^{i\phi}  
       \end{array} \right) \!\!
\left( \begin{array}{cc}
       1 & 0 \\
       0 & i\, e^{-i\psi + i\chi}  
       \end{array} \right) .  
\] 
For more details about the case U(2), the reader is referred to Appendix~C.

No analytic solution is found as soon as $n>2$.
Even the decomposition of an arbitrary member of U(3) is an unsolved 
problem\footnote{As soon as one of the nine entries of the U(3) matrix equals zero,
                 the problem is analytically solvable.},
in spite of substantial efforts by the authors of the present paper.
Independently, in the framework of an other but related problem,
Shchesnovich \cite{shchesnovich} comes to a similar conclusion.
For arbitrary~$n$, Reference \cite{bremen} gives an analytic solution 
for a $2n$-dimensional subset of the $n^2$-dimensional group~U($n$).
We conjecture that the asymptotic scaling procedure of Sections~1 and~2 provides
a numerical solution, for any member of U($n$), with arbitrary~$n$. 
  
If, in particular, we have $n=2^w$, then a U($n$) matrix represents
a quantum circuit of width~$w$, i.e.\ acting on $w$~qubits.
We thus may conclude that such circuit can be decomposed as the cascade 
of an overall phase, two $Z$ subcircuits and one $X$ subcircuit.
The basic building block of any $Z$ circuit is the 1-qubit circuit represented by $2 \times 2$ matrix
\[
{\tt PHASOR}(\teta) = \left( \begin{array}{cc} 1 & 0 \\ 0 & e^{i\teta}\end{array} \right) \ ; 
\]
the basic building block of any $X$ circuit is the 1-qubit circuit represented by 
\[
{\tt NEGATOR}(\teta) = \left( \begin{array}{cc} \cos(\teta)e^{-i\teta} & i\sin(\teta)e^{-i\teta} \\
                                               i\sin(\teta)e^{-i\teta} &  \cos(\teta)e^{-i\teta} \end{array} \right) \ . 
\]
The {\tt NEGATOR} realizes an arbitrary root of {\tt NOT} \cite{osid} \cite{victoria1} \cite{brussel}
and thus is a natural generalization of the square root of the {\tt NOT} gate \cite{debeule}.

Each $2^w \times 2^w$ matrix $Z$ is implemented by a string of $2^{w-1}$ controlled {\tt PHASOR}s; 
any  $2^w \times 2^w$ matrix~$X$ represents a circuit composed of controlled {\tt NEGATOR}s \cite{osid}. 

By noting the identity
\[
\mbox{diag}( a, a, a, a, a, ..., a, a) = \ X_0 \ 
\mbox{diag}( 1, a, 1, a, 1, ..., 1, a) \ X_0^{-1} \ 
\mbox{diag}( 1, a, 1, a, 1, ..., 1, a) \ ,
\]
where $a$ is a short-hand notation for $e^{i\alfa}$ and $X_0$ is the permutation matrix
\[
 \left( \begin{array}{ccccccc} 0     & 1 & 0 & 0 & ... & 0 & 0 \\
                               0     & 0 & 1 & 0 & ... & 0 & 0 \\
                               0     & 0 & 0 & 1 & ... & 0 & 0 \\
                              \vdots &   &   &   &     &   &   \\
                               0     & 0 & 0 & 0 & ... & 0 & 1 \\
                               1     & 0 & 0 & 0 & ... & 0 & 0 \end{array} \right) \ ,
\]
we can transform (\ref{ZXZ}) into a decomposition containing exclusively XU and ZU matrices:
\begin{equation}
U = X_0Z_0X_0^{-1}Z_1'XZ_2 \ ,
\label{XZXZXZ}
\end{equation}
where $X_0$ is an XU matrix which can be implemented with classical reversible gates 
(i.e.\ {\tt NOT}s and controlled {\tt NOT}s), 
where $Z_0$ is a ZU matrix which can be implemented by a single (uncontrolled) {\tt PHASOR} gate, and
where $Z_1'$ is the product $\mbox{diag}( 1, a, 1, a, 1, ..., 1, a)\, Z_1$.
The short decomposition (\ref{XZXZXZ}) illustrates the power of the two subgroups
XU($n$) and ZU($n$), which are complementary \cite{acm} \cite{bremen},
in the sense that
\begin{itemize}
\item they overlap very little, as their intersection is the trivial group 
      consisting of merely the $n \times n$ unit matrix and
\item they strengthen each other sufficiently, as their closure is the whole unitary group U($n$).
\end{itemize}
As an example, we give here a decomposition of the Hadamard gate according to
schemes (\ref{ZXZ}) and (\ref{XZXZXZ}), respectively:
\bea
\hspace*{-20mm}
& &
\frac{1}{\sqrt{2}}\ 
\left( \begin{array}{rr}  1             &   1           \\      1             & -1                     \end{array} \right)\ = \  \begin{array}{r} \frac{1+i}{\sqrt{2}} \end{array}    
\left( \begin{array}{rr}  1             &   0           \\      0             &  i                     \end{array} \right)
\left( \begin{array}{rr}  \frac{1-i}{2} & \frac{1+i}{2} \\[1mm] \frac{1+i}{2} &  \frac{1-i}{2}         \end{array} \right)
\left( \begin{array}{rc}  1             &   0           \\      0             & -i                     \end{array} \right) \\[1.5mm] & = &
{\footnotesize 
\left( \begin{array}{rr}  0             &   1           \\      1             &  0                     \end{array} \right)
\left( \begin{array}{rc}  1             &   0           \\      0             &  \frac{1+i}{\sqrt{2}}  \end{array} \right)
\left( \begin{array}{rr}  0             &   1           \\      1             &  0                     \end{array} \right)
\left( \begin{array}{rc}  1             &   0           \\      0             &  \frac{-1+i}{\sqrt{2}} \end{array} \right)
\left( \begin{array}{rr}  \frac{1-i}{2} & \frac{1+i}{2} \\[1mm] \frac{1+i}{2} &  \frac{1-i}{2}         \end{array} \right)
\left( \begin{array}{rr}  1             &   0           \\      0             & -i                     \end{array} \right)} \ .
\eea    

\section{Conclusion}

We have presented a method for scaling an arbitrary matrix from the unitary group U($n$),
by multiplying the matrix to the left and to the right by unitary diagonal matrices.
We conjecture that the resulting scaled matrix is a member of XU($n$),
i.e.\ the subgroup of U($n$) consisting of all $n \times n$ unitary matrices
with all $2n$~line sums equal to~1.
If $n=2$, then scaling can be performed analytically and thus with infinite precision.
If $n>2$, then scaling has to be performed numerically and thus with finite precision.
In the terminology of Linial et al.~\cite{linial}, we would say that
matrices from U(2) are `scalable', whereas 
matrices from U($n$) with $n>2$ are `almost scalable'. 
The conjecture that the numerical algorithm converges to a unit line-sum matrix,
is based on four observations:
\begin{itemize}
\item the proof that such matrix exists in the case $n=2$,
\vspace*{-1mm}
\item the proof that such matrix exists for a $2n$-dimensional subset of the general case~U($n$) \cite{bremen},
\vspace*{-1mm}
\item the success of 1,000 numerical experiments in the cases $n=3$, $n=4$, and $n=5$, and
\vspace*{-1mm}
\item the fact that, according to (\ref{dim}), there are, for arbitrary~$n$,  
      exactly the right number of freedoms.
\end{itemize}

For the special case of $n=2^w$,
this leads to a decomposition of an arbitrary quantum circuit into
\begin{itemize} 
\item one overall phase, one X~circuit and two Z~circuits or
\vspace*{-1mm}
\item three X~circuits and three Z~circuits.
\end{itemize}


\appendix

\section{Theorem}

Theorem : 
The absolute value of the matrix sum of a U($n$) matrix 
is smaller than or equal to~$n$. 
The U($n$) matrices with abs(matrixsum)~=~$n$
are member of the subgroup $e^{i\alfa}$~XU($n$),
where XU($n$) denotes the subgroup of U($n$) 
consisting of the matrices with unit line sums.

\subsection*{Proof}

Let $r_1$, $r_2$, ..., and $r_n$ be the row sums of an $n \times n$ matrix.
For convenience, we give their real and imaginary parts an explicit notation:
\[
r_j = s_j + i t_j \ .
\]
The matrix sum is
\[
m = r_1 + r_2 + ... + r_n \ .
\]
If the matrix is unitary, then we have
\[
|r_1|^2 + |r_2|^2 + ... + |r_n|^2 = n \ ,
\]
as proved in Appendix~A of Reference~\cite{vanlaer}.
We rewrite this property as follows:
\begin{eqnarray}
s_1^2 + s_2^2 + ... + s_n^2 & = & n\,\cos^2(\Sigma) \label{s^2} \\[1mm] 
t_1^2 + t_2^2 + ... + t_n^2 & = & n\,\sin^2(\Sigma) \nonumber \ , 
\end{eqnarray}
where the angle $\Sigma$ is allowed to have any value.

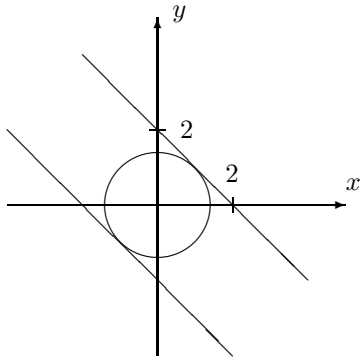
\begin{figure} 
\centering
\setlength{\unitlength}{10mm}
\begin{picture}(2,2)(-1,-1)
\put(-2,0){\vector(1,0){4.5}}
\put(0,-2){\vector(0,1){4.5}}
\put(2.5,0.2){$x$}
\put(0.2,2.5){$y$}
\put(1,-0.1){\line(0,1){0.2}}
\put(0.9,0.3){2}
\put(-0.1,1){\line(1,0){0.2}}
\put(0.3,0.9){2}
\put(0,0){\circle{10}}
\put(-1,2){\line(1,-1){3}}
\put(-2,1){\line(1,-1){3}}
\end{picture}
\vspace*{10mm}
\caption{The circle $x^2+y^2=2$ and the set of two straight lines $(x+y)^2=4$,
         with tangent points (1,~1) and ($-1$,~$-1)$.}
\label{cirkel+2rechten}
\end{figure}

We consider (\ref{s^2}) as the eqn of an $n$-dimensional hypersphere.
We ask ourselves what is the highest value of the function
\[
f(s_1, s_2, ..., s_n) = (s_1 + s_2 + ... + s_n)^2
\]
on the surface of this hypersphere.
For this purpose, we note that $f = k$, with $k$ some positive constant,
is the eqn of the set of two parallel hyperplanes:
\[
s_1 + s_2 + ... + s_n = \pm\, \sqrt{k} \ .
\]
The highest value of $k$ on the hypersphere is when the two planes
are tangent to the sphere. This happens when $k$ equals $n^2\,\cos^2(\Sigma)$,
the two tangent points having coordinates 
$(s_1, s_2, ..., s_n) = \pm \cos(\Sigma)\ ( 1, 1, ..., 1)$
and $f$ having the value $n^2\cos^2(\Sigma)$. 
See the 2-dimensional illustration in Figure~\ref{cirkel+2rechten}.

A similar reasoning is possible for the function
\[
g(t_1, t_2, ..., t_n) = (t_1 + t_2 + ... + t_n)^2 \ .
\]
Noting that $|m|^2$ equals $f+g$, we conclude that $|m|^2$ has maximum value
$n^2\,\cos^2(\Sigma) + n^2\,\sin^2(\Sigma) = n^2$.
The unitary matrices with this particular $|m|^2$ value are the matrices with
$r_j = \pm \cos(\Sigma) \pm i \sin(\Sigma)$,
i.e.\ the matrices with constant row sum equal to $e^{i\alfa}$,
where $\alfa$ is either $\Sigma$ or $\Sigma + \pi$.

A dual reasoning holds for the column sums,
with $c_j = d_j + ie_j$, such that 
$d_1^2 + d_2^2 + ... + d_n^2 = n \cos^2(\Delta)$ and 
$e_1^2 + e_2^2 + ... + e_n^2 = n \sin^2(\Delta)$.
The unitary matrices with $|m|^2 = n^2$ are the matrices with
$c_j = \pm \cos(\Delta) \pm i \sin(\Delta)$,
i.e.\ the matrices with constant column sum equal to $e^{i\beta}$,
where $\beta$ is either $\Delta$ or $\Delta + \pi$.
Because a matrix can have only one matrix sum,
a matrix with both constant row sum and constant column sum
necessarily has constant line sum. 
Therefore, for the matrices with $|m|^2 = n^2$, the angle $\beta$ equals the angle $\alfa$.

The maximum-$|m|$ matrices equal $e^{i\alfa}$ times 
a matrix with constant line sum equal to~1.
Thus they are member of the group $e^{i\alfa}$~XU($n$), a~subgroup of U($n$), 
isomorphic to U(1)~$\times$~XU($n$), and thus isomorphic to U(1)~$\times$~U($n-1$).

\section{The potential landscape}


Given a unitary matrix~$A$,
finding the scaled matrix $B$ is equivalent to solving the (non-linear) eqn
\[
n^2 - |\som(B)|^2 = 0 \ .
\]
As 
\begin{equation}
B_{jk} = e^{i(\lambda_j + \rho_k)}\, A_{jk}\ ,
\label{B}
\end{equation}
we introduce a $2n$-dimensional landscape $\Psi$, given by
\[
\Psi(\lambda_1, \lambda_2, ..., \lambda_n, \rho_1, \rho_2, ..., \rho_n) = n^2-|\som(B)|^2 \ .
\]
We have to find the minimum of this function, i.e.\ the point $\Psi=0$.

In order to investigate the shape of the $\Psi$~function, we linearize the equation
around $A$, i.e.\ in the neighbourhood of  
$(\lambda_1, \lambda_2, ..., \lambda_n, \rho_1, \rho_2, ..., \rho_n)$~= $(0, 0, ..., 0,$ $0, 0, ...,0)$.
For this purpose, we write the row sums, column sums, and matrix sum 
of the given matrix~$A$ as follows:
\bea
r_j & = & s_j + it_j \\
c_j & = & d_j + ie_j \\
m   & = & p   + iq   \ .
\eea
From (\ref{B}) we deduce
\bea
{\cal R}e [\som(B)] & \approx & p - \sum_j t_j\lambda_j - \sum_j e_j\rho_j \\ 
{\cal I}m [\som(B)] & \approx & q + \sum_j s_j\lambda_j + \sum_j d_j\rho_j \ ,
\eea
such that
\[
\Psi \approx n^2 - p^2 - q^2 + 2\ \sum_j(pt_j - qs_j)\, \lambda_j + 2\ \sum_j(pe_j - qd_j)\, \rho_j \ .
\]
The coefficients of $\lambda_j$ and $\rho_j$ form 
the gradient vector of the $\Psi$ landscape.
A~stationary point occurs whenever, for all~$j$,
\bea
pt_j - qs_j & = & 0 \\
pe_j - qd_j & = & 0 \ .
\eea
These conditions can only be fulfilled in the following cases:
\begin{itemize}
\item when \[ p=q=0 \ , \]
      i.e.\ when the matrix sum is zero and hence $\Psi$ has the global maximum value of $n^2$,
\item when \[  \mbox{ all } t_j=e_j=0 \ , \]
      i.e.\ when all line sums are real,
\item when \[  \mbox{ all } s_j=d_j=0 \ , \]
      i.e.\ when all line sums are imaginary,
\item or when 
      \[ \mbox{ either } s_j=t_j=0 \mbox{ or } \frac{s_j}{t_j} = \frac{p}{q} \]
      together with 
      \[ \mbox{ either } d_j=e_j=0 \mbox{ or } \frac{d_j}{e_j} = \frac{p}{q} \ , \]
      i.e.\ when all non-zero line sums have the same argument.
\end{itemize}
We conjecture that all these stationary points are either maxima or saddle points or global minima. 
In other words: we conjecture that no local minima exist.
Moreover, we conjecture that the global minima satisfy~$\Psi=0$.

\section{The case U(2)}

We consider the unitary group U(2).
All $2 \times 2$ diagonal unitary matrices form the subgroup DU(2),
isomorphic to U(1)~$\times$~U(1).
The subgroup DU(2) divides its supergroup U(2) into double cosets.
Let $A$ be an arbitrary U(2) matrix:
\[
A = U(\phi, \teta, \psi, \chi) = 
e^{i\teta}
\left( \begin{array}{rl}
       \cos(\phi)\, e^{ i\psi}\  & \sin(\phi)\, e^{ i\chi} \\
      -\sin(\phi)\, e^{-i\chi}   & \cos(\phi)\, e^{-i\psi}  
       \end{array} \right) \ .
\]
Its double coset consists of all matrices
\[
\left( \begin{array}{cc}
       e^{i\lambda_1} & 0 \\
       0 & e^{i\lambda_2}  
       \end{array} \right) 
\, A\, 
\left( \begin{array}{cc}
       e^{i\ro_1} & 0 \\
       0 & e^{i\ro_2}  
       \end{array} \right) \ =
\left( \begin{array}{rr}
       c\, e^{i(\teta+\psi+\lambda_1+\ro_1)} & s\, e^{i(\teta+\chi+\lambda_1+\ro_2)} \\
      -s\, e^{i(\teta-\chi+\lambda_2+\ro_1)} & c\, e^{i(\teta-\psi+\lambda_2+\ro_2)}  
       \end{array} \right) \ ,
\]
where $c$ and $s$ are short-hand notations for $\cos(\phi)$ and $\sin(\phi)$, respectively.
We introduce the variables
\bea
u & = & \lambda_1 + \ro_1 \\ 
v & = & \lambda_1 + \ro_2 \\ 
w & = & \lambda_2 + \ro_1 \\ 
t & = & \lambda_2 + \ro_2
\eea
and note the identity
\[
u - v - w + t = 0 \ .
\]
Therefore, the double coset is the 3-parameter space   
\[
\left( \begin{array}{rl}
       c\, e^{i(\teta+\psi+u)}  & s\, e^{i(\teta+\chi+v    )} \\
      -s\, e^{i(\teta-\chi+w)}  & c\, e^{i(\teta-\psi-u+v+w)}  
       \end{array} \right) \ .
\]
For convenience, we change variables:
\bea
x & = & \frac{1}{2}\, v + \frac{1}{2}\, w + \teta \\
y & = & u - \frac{1}{2}\, v - \frac{1}{2}\, w + \psi \\
z & = & \frac{1}{2}\, v - \frac{1}{2}\, w + \chi \ ,
\eea
resulting in
\[
e^{ix}\ 
\left( \begin{array}{rl}
       c\, e^{iy}\  & s\, e^{iz} \\
      -s\, e^{-iz}  & c\, e^{-iy}  
       \end{array} \right) \ .
\]
Thus the double coset of $A$ consists of the matrices $U(\phi, x, y, z)$,
i.e.\ of all matrices with the same value of the angle~$\phi$.
This constitutes a 3-dimensional subspace of the 4-dimensional space~U(2),
except for the cases $s=0$ 
(i.e.\ for the double coset of the {\tt IDENTITY} matrix 
${\tiny \left( \begin{array}{cc} 1 & 0 \\ 0 & 1 \end{array} \right)}$) and $c=0$ 
(i.e.\ for the double coset of the {\tt NOT}      matrix 
${\tiny \left( \begin{array}{cc} 0 & 1 \\ 1 & 0 \end{array} \right)}$),
which both are 2-dimensional 
only\footnote{One may consider an arbitrary place $P(\varphi, \lambda)$ on earth.
              The points $Q(\varphi, x)$, 
              with same latitude~$\varphi$ but arbitrary longitude~$x$, 
              form a 1-dimensional subspace 
              of the 2-dimensional earth surface, called the parallel of~$P$, except if 
              either $\varphi= \pi/2$,
              in which case $Q(\varphi, x)$ is a 0-dimensional subspace, called the north pole,
              or     $\varphi=-\pi/2$,
              in which case $Q(\varphi, x)$ is a 0-dimensional subspace, called the south pole.
              We therefore may consider 
              the 2-dimensional double coset of the {\tt IDENTITY} matrix and 
              the 2-dimensional double coset of the {\tt NOT}      matrix
              as the north and the south pole, respectively, of the 4-dimensional U(2) manifold.}.
             
What are, within the double coset of~$A$, the stationary points of the $\Psi$ landscape?
We easily find 
\[
\Psi(x,y,z) = 4 - 4\,[\, c^2\cos^2(y) + s^2\sin^2(z)\, ] \ .
\]
The conditions 
$\partial \Psi/\partial x = 0$, 
$\partial \Psi/\partial y = 0$, and 
$\partial \Psi/\partial z = 0$ 
immediately lead to
\bea
\sin(2y) & = & 0 \\
\sin(2z) & = & 0 \ . 
\eea
This set of two trigonometric equations in the three unknowns $x$, $y$, and~$z$
has infinitely many solutions, leading to an infinite number of matrices:
\[
U(\phi, x, k\,\frac{\pi}{2}, l\,\frac{\pi}{2})
\]
with $x$ arbitrary, $k \in \{ 0, 1, 2, 3\}$, and $l \in \{ 0, 1, 2, 3\}$.
These sixteen sets of matrices lead to $\Psi$ values equal to $4$, $4c^2$, $4s^2$, and $0$,
corresponding to global maxima, saddle points, saddle points, and global minima,
respectively.
The saddle points belong to U(1)$\times$O(2), subgroup of U(2);
the global extrema do~not.

What are, within the same double coset, the matrices with all line sums real? 
We easily find the conditions:
\bea
 c\,\sin(x+y) + s\,\sin(y+z) & = & 0 \\
-s\,\sin(x-z) + c\,\sin(x-y) & = & 0 \\
 c\,\sin(x+y) - s\,\sin(x-z) & = & 0 \ , 
\eea
leading to 
\bea
   \sin(x-y) & = & \ \ \, \sin(x+y) \\
   \sin(x-z) & = & -      \sin(y+z) \\
s\,\sin(x-z) & = & c\,    \sin(x+y) \ . 
\eea
This set of three trigonometric equations in the three unknowns $x$, $y$, and $z$
has twelve solutions:
\bea
&& U(\phi, 0,   0, 0),             \\
&& U(\phi, 0, \pi, 0),             \\
&& U(\phi, \pi/2, -\pi/2, -\pi/2), \\
&& U(\phi, \pi/2, -\pi/2,  \pi/2), \\
&& U(\phi, \pi/2,  \pi/2, -\pi/2), \\
&& U(\phi, \pi/2,  \pi/2,  \pi/2), \\
&& U(\phi, \pi,   0, 0),           \\
&& U(\phi, \pi, \pi, 0),           \\
&& U(\phi, -\phi,   0,  \pi/2),    \\
&& U(\phi, -\phi, \pi, -\pi/2),    \\
&& U(\phi,  \phi,   0, -\pi/2), \mbox{ and } \\
&& U(\phi,  \phi, \pi,  \pi/2)    \ .
\eea
Four matrices have matrix sum $0$ and thus $\Psi = 4$;
four matrices have matrix sum $\pm 2s$ and thus $\Psi = 4\cos^2(\phi)$;
four matrices have matrix sum $\pm 2c$ and thus $\Psi = 4\sin^2(\phi)$;
four matrices have matrix sum $\pm 2$  and thus $\Psi = 0$.
Thus 
four  of the twelve matrices represent a global maximum;
eight of the twelve matrices represent a saddle point;
four  of the twelve matrices represent a global minimum.

As an example, we choose the $A$~matrix with $0 < \phi < \pi/4$, such that $0 < s < c < 1/\sqrt{2}$.
Among the twelve matrices of its double coset with real line sums , 
there are only four matrices where the four line sums are positive, i.e.\
\bea
S   & = & U(\phi,     0,   0,      0) =              \left( \begin{array}{rr} c &   s \\ - s &  c \end{array} \right) ,                \\[1mm]  
S'  & = & U(\phi,   \pi, \pi,      0) =              \left( \begin{array}{rr} c & - s \\   s &  c \end{array} \right) = S^{\, -1} ,    \\[1mm] 
B   & = & U(\phi, -\phi,   0,  \pi/2) = \frac{1}{e}\ \left( \begin{array}{rr} c &  is \\  is &  c \end{array} \right) \ , \mbox{ and } \\[1mm] 
B'  & = & U(\phi,  \phi,   0, -\pi/2) =          e \ \left( \begin{array}{rr} c & -is \\ -is &  c \end{array} \right) = B^{\, -1} , 
\eea
where $e$ is a short-hand notation for $e^{i\phi}=c+is$.
Among these four matrices, only $B$ and $B'$ have unit line sum (and thus $\Psi=0$).

In the neighbourhood of $S$, we have the matrices
\[
(1+ix-x^2/2)\ 
\left( \begin{array}{rl}  c\, (1+iy-y^2/2) &  s\, (1+iz-z^2/2) \\ 
                         -s\, (1-iz-z^2/2) &  c\, (1-iy-y^2/2) \end{array} \right)
\]
with $x$, $y$, and $z$ small.
This yields a matrix sum
\[
2c - c\, (x^2 + y^2) - 2sxz + 2i\, (cx+sz) 
\]
and thus a potential
\[
\Psi(x, y, z) = 4s^2 + 4c^2y^2 - 4s^2z^2 \ .
\]
The opposite signs of the coefficients of $y^2$ and $z^2$ 
illustrate the fact that $S$ is a saddle point of the $\Psi$ landscape.
Only if the subsequent matrices $A_k, A_{k+1}, ...$ are situated on the $z=0$~line,
then the Sinkhorn-like procedure of Section~2 halts at the point $S$.
In order to leave this stop, it suffices to continue along another line, e.g.\ $y=0$.
Similar conclusions hold for the point~$S'$.

In the neighbourhood of $B$, we have the matrices
\[
\frac{1}{e}\ (1+ix-x^2/2)\ 
\left( \begin{array}{ll}  c\  (1+iy-y^2/2) & is\, (1+iz-z^2/2) \\ 
                         is\, (1-iz-z^2/2) &  c\  (1-iy-y^2/2) \end{array} \right)
\]
with $x$, $y$, and $z$ small.
This yields a matrix sum
\[
2 - x^2 - c^2y^2 - s^2z^2 + i\, (2x + scy^2 - scz^2)
\]
and thus a potential
\begin{equation}
\Psi = 4c^2y^2 + 4s^2z^2 \ .  
\label{X^2+Y^2}
\end{equation}
The positive signs of the coefficients of $y^2$ and $z^2$ 
illustrate the fact that $B$ is 
a minimum (actually, a global minimum) of the $\Psi$ landscape.
The same conclusion holds for the point~$B'$.

Let us assume that, in spite of the direct analytic solution for $n=2$,
we scale a~U(2) matrix by the iterative method of Sections~1 and~2.
Once close to the point~$B$, how fast do we converge to this global minimum of~$\Psi$?
Close to~$B$, we have $A_k$ of the form
\begin{equation} 
\frac{1}{e}\ (1+ix) \ 
\left( \begin{array}{ll}  c\  (1+iy) & is\, (1+iz) \\ 
                         is\, (1-iz) &  c\  (1-iy) \end{array} \right) \ .
\label{3D}
\end{equation}
As soon as $k>0$, both column sums of $A_k$ are real (or zero), 
such that $x=0$ and $z=(c^2/s^2)y$. Thus we have a matrix
\begin{equation}
A_k = \frac{1}{e}\   
\left( \begin{array}{ll}  c\  (1+iy)        & is\, (1+ic^2y/s^2)      \\ 
                         is\, (1-ic^2y/s^2) &  c\  (1-iy) \end{array} \right) \ ,
\label{1D}
\end{equation}
and, because of (\ref{X^2+Y^2}), a potential $\Psi(A_k)=(4c^2/s^2)\, y^2$.
Thus all matrices $A_k$ lay on a line, the 1-dimensional space~(\ref{1D}),
subspace of the 3-dimensional space~(\ref{3D}). 
If we apply the $(k+1)$~th step of the iterative algorithm, we find, after some algebra:
\[
A_{k+1} = \frac{1}{e}\  
\left( \begin{array}{ll}  c\  (1+iay)        & is\, (1+ic^2ay/s^2)      \\ 
                         is\, (1-ic^2ay/s^2) &  c\  (1-iay) \end{array} \right) \ ,
\]
where $a=1-4c^2s^2=\cos^2(2\phi)$. 
Hence the new potential is $\Psi(A_{k+1})=(4c^2/s^2)\, (ay)^2$ and
\[
\frac{\Psi(A_{k+1})}{\Psi(A_k)} =\cos^4(2\phi) \ .
\]
This illustrates the fact that the convergence speed of the algorithm is indeed 
dependent on the given matrix~$A$, more specifically on its parameter~$\phi$.
If this angle is close to $\pi/4$, then convergence is fast;
if the  angle is close to     $0$, then convergence is slow.

Finally, we ask ourselves, given the matrix~$A$,
does the algorithm of Sections~1 and~2 lead 
to the scaled matrix~$B$ or 
to the scaled matrix~$B'$~?
The separatrice consists of the spaces
$\chi=\psi$ and $\chi=\psi+\pi$.
If $0     < \chi-\psi < \pi$, then the trajectory $A_0$, $A_1$, $A_2$,~... ends 
in the attractor $B$; 
if $-\pi  < \chi-\psi < 0$,   then the trajectory $A_0$, $A_1$, $A_2$,~... ends 
in the attractor $B'$;
if $\chi-\psi = 0$ or $\chi-\psi = \pi$, 
then $A_1$ is an orthogonal matrix (either $S$ or $S'$) and
thus a saddle-point, such that the final destination (either $B$ or $B'$)
depends on the direction in which one leaves the saddle point. 

We close this appendix by comparing 
the above quantitative U(2) results with the qualitative U($n$) properties.
It is well-known that,
if a finite group G has a subgroup H, then H divides G into double cosets
with sizes ranging from order(H) to order$^2$(H).
Similarly, if a Lie group G has a Lie subgroup H, then H divides G into double cosets
with dimension ranging from dim(H) to 2$\,$dim(H).
The group U($n$) is $n^2$-dimensional and its subgroup DU($n$) is $n$-dimensional.
As a result, DU($n$) divides U($n$) into double 
cosets\footnote{This set of double cosets, i.e.\ the double coset space
                \[ \mbox{U}(1)^n \setminus \mbox{U}(n)\, /\ \mbox{U}(1)^n \]
                can be mapped to the set (not group!) of so-called
                unistochastic $n \times n$ matrices \cite{bengtsson},
                a~subset of the well-known semigroup of 
                $n \times n$ bistochastic matrices
                (a.k.a.\ doubly stochastic matrices).},
each with dimension between $n$ and $2n$.
In fact, in this particular case, the dimensions of the double cosets range
from $n$ to $2n-1$.
Most of the double cosets are $(2n-1)$-dimensional;
only some are lower-dimensional, e.g.\ the $n!$ double cosets of 
permutation matrices being only 
$n$-dimensional\footnote{Together these $n!$~double cosets form the group of 
                         complex permutation matrices, 
                         a group isomorphic to the semidirect product
                         DU($n$)~{\bf :}~S$_n$,
                         where S$_n$ is the symmetric group of degree~$n$.}.
Thus within a double coset, we have at most $2n-1$ degrees of freedom.
If we want a matrix with all line sums real,
then this imposes $2n-1$ conditions, usually lowering the number of freedoms to~0.
In other words: 
in each double coset there usually are a finite number of real line-sum matrices.
We conjecture that at least one of these matrices is a unit line-sum matrix.

\end{document}